\documentclass[11pt,a4paper]{article}
\pdfoutput=1
\usepackage{jcappub}

\usepackage{color}
\input{colordvi.tex}

\usepackage{amsmath}
\usepackage{float}
\usepackage{wrapfig}
\usepackage{bm}
\usepackage{graphicx}
\usepackage{subfigure}
\usepackage{verbatim}
\usepackage{makeidx}
\usepackage{amssymb}
  
\newcommand{\vect}[1]{\mbox{\boldmath${#1}$}}

\title{Probing small-scale non-Gaussianity from anisotropies in
acoustic reheating}
\author{Atsushi Naruko,} 
\author{Atsuhisa Ota}
\author{and Masahide Yamaguchi} 
\affiliation{Department of Physics, Tokyo Institute of Technology, \\
Tokyo 152-8551, Japan}
\emailAdd{a.ota@th.phys.titech.ac.jp}

\abstract{We give new constraints on small-scale non-Gaussianity of
primordial curvature perturbations by the use of anisotropies in
acoustic reheating. Mixing of local thermal or local kinetic equilibrium
systems with different temperatures yields a locally averaged temperature
rise, which is proportional to the square of temperature perturbations
damping in the photon diffusion scale. Such secondary temperature
perturbations are indistinguishable from the standard temperature
perturbations linearly coming from primordial curvature perturbations
and hence should be subdominant compared to the standard ones. We show
that small-scale higher order correlation functions (connected
non-Gaussian and disconnected Gaussian parts) of primordial curvature
perturbations can be probed by investigating auto power spectrum of the
generated secondary perturbations and the cross power spectrum with the
standard perturbations. This is simply because these power spectra come
from higher order correlation functions of primordial curvature
perturbations with non-linear parameters such as $f_{\rm NL}$ and
$\tau_{\rm NL}$ since secondary temperature perturbations are second
order effects.
Thus, the observational results $l(l+1)C^{TT}_l\simeq 6\times 10^{-10}$
at large scales give a robust and universal upper bound on small-scale
non-Gaussianities of primordial curvature perturbations.}


\begin{document}
\maketitle

\section{Introduction}

The 21st century is the age of precise observational cosmology.  It is
possible to observationally test theoretical models for the early
universe by analysing anisotropies of the Cosmic Microwave Background
(CMB)~\cite{Hinshaw:2012aka,Ade:2015lrj}.  In fact, we already
know that primordial curvature perturbations do exist, and their
spectrum is almost scale-invariant.  While the statistical property of
primordial perturbations is well approximated by Gaussian statistics,
the deviation from the exact Gaussian statistics~\cite{Komatsu:2001rj}
will deliver us rich information on the primordial
universe~\cite{Maldacena:2002vr}, which is one of the next targets in
cosmology.  On the other hand, it should be noticed that the scales
probed by the CMB anisotropies are just $6$ $e$-foldings of the last
$60$ $e$-folds during inflation and that we know very little about
smaller-scale perturbations on $k~{\rm Mpc}\gtrsim 0.1$ because most of
the primordial density fluctuations dissipate due to Silk damping. One
powerful tool to see the fluctuations on small scales is CMB spectral
distortions~\cite{Zeldovich:1969ff,Sunyaev:1970er,Hu:1994bz,1991MNRAS.248...52B,1991ApJ...371...14D,Chluba:2012we,Khatri:2012rt,Khatri:2012tv,Chluba:2012gq}.
Deviations from the ideal blackbody spectrum are induced in the early
universe after around $z \simeq 2 \times 10^6$,
and such deviations are characterized in terms of several parameters
such as the Compton $y$ parameter and the chemical potential $\mu$.  The
next observational projects, such as PIXIE and
PRISM~\cite{Kogut:2011xw,Andre:2013afa}, following COBE/FIRAS are
proposed~\cite{Mather:1993ij,Fixsen:1996nj,Salvaterra:2002mg}, and it
may be possible to see the nature of primordial fluctuations up to
another $11$ $e$-foldings in the near future.  Hence, $17$ $e$-foldings
will be within the scope of our current technology, but more than one half
of total $e$-foldings necessary to solve the initial condition problems
remain obscure.

Recently, {\it acoustic reheating} was proposed to investigate
primordial curvature perturbations on extremely small
scales~\cite{Jeong:2014gna,Nakama:2014vla}.  Below the diffusion scale,
the inhomogeneities are erased due to Silk damping so that the universe
is heated by the conversion of the dissipated energy of
photon~\cite{Chluba:2004cn}. Such temperature rises are quadratic in
temperature fluctuations over the diffusion scale. In the previous
works, the total average (homogeneous part) of the temperature rise was
discussed, and they derived the constraints on small-scale power
spectrum of primordial curvature perturbations by comparing the helium
mass fraction ${\rm Y}_p$ (or photon baryon ratio $\eta_{B}$)
during Big Bang nucleosynthesis with that at the last scattering of
the CMB. Such investigations are new and interesting, but the scales
probed by these approaches are limited up to $k{\rm Mpc}\simeq \mathcal
O(10^5)$ since the heating in much smaller scales (at much earlier
epochs) changes both $\eta^{\rm BBN}_B$ and $\eta^{\rm CMB}_B$, so that
we know very little about the scale except the constraints that
$\mathcal P_{\mathcal R}\lesssim 0.3$ for $k\sim 10^{20-25}{\rm
Mpc}^{-1}$~\cite{Jeong:2014gna}. However, the temperature rise induced
by acoustic reheating is only homogenized over the diffusion scale.  Then,
it fluctuates spatially in general and potentially includes all of the
histories of erased temperature perturbations from the beginning of the
hot universe, that is, if large inhomogeneities would exist at small
scales, the observed temperature anisotropies can be changed
drastically. In this paper, we shall focus on such inhomogeneities of
acoustic reheating. Although one cannot pick up and isolate the
temperature perturbations due to acoustic reheating from the observed
temperature anisotropies, given the fact that the linear theory
prediction well matches with the CMB observations, the non-linear
corrections should not be beyond the observational $C^{TT}_l$, i.e. the
order of $10^{-10}$ at large scales. This puts robust and universal
constraints on higher order correlation functions including
non-Gaussianities of primordial curvature perturbations since the
non-linear temperature corrections are of the second order of primordial
curvature perturbations. 
Our approach is closely related with a method
to investigate anisotropies in CMB distortions which was originally
proposed by Pajer and Zaldarriaga since the physics of acoustic
reheating is essentially the same~\cite{Pajer:2012vz}.

We organize this paper as follows.  In section \ref{section2}, we
summarize the details of acoustic reheating, particularly paying
attention to its perturbations. Section \ref{section3} and
\ref{section4} are devoted to the calculations of angular power spectrum
by the use of Sachs-Wolfe approximation, and we obtain the constraints
on the non-Gaussianities of primordial curvature perturbations in
section \ref{section5}. We give conclusions and discussions in the final
section, and scale-dependence of non-Gaussianities is briefly mentioned
in the appendix.

\section{Inhomogeneities in acoustic reheating}\label{section2}

The local photon temperature in the early universe fluctuates due to the
primordial curvature perturbations. As long as the Compton and the
double-Compton processes are efficient, which holds for $z \gtrsim
2\times 10^6$~\cite{Danese:1982,Chluba:2006kg}, it is locally in thermal
equilibrium, and its spectrum obeys the Planck distribution
characterized only by the local temperature $T(\eta,\vect{x},\hat{\vect
n})$, where $\eta$, $\vect x$ and $\hat{\vect n}$ are the conformal
time, the space coordinate and the unit vector of the photon momentum
respectively. This local temperature can be divided into the homogeneous
part and the inhomogeneous one,
\begin{align}
T(\eta,\vect x, \hat{\vect n}) = \overline{T}(\eta) \bigl( 1 + \Theta(\eta, \vect x,\hat{\vect n}) \bigr).\label{Tdef}
\end{align}
Here, the acoustic reheating is not yet taken into account, and hence
$\overline{T}$ decreases in proportional to $a(\eta)^{-1}$ ($a(\eta)$ :
the scale factor) in the expanding Universe as long as the number of the
effective degree of freedom of radiations is unchanged, and the
temperature perturbations are proportional to the primordial curvature
perturbations at linear order.  Once the acoustic reheating happens,
small-scale inhomogeneities are erased due to Silk damping, and the
effective temperature is raised because of the energy conservation. Now,
let us evaluate the temperature rise for a given perturbed system of
photons, assuming acoustic reheating happens instantaneously. Keeping in
mind the relation between the temperature and the energy density of
photons $\rho$, $T=(\rho/a_B)^{1/4}$ with the numerical constant
$a_B=\pi^2/15$, the dimensionless temperature rise $\Delta$ is given by
\begin{align}
  \Delta(\vect x) = \frac{\langle \rho \rangle^{\frac14}_{\bm x}-\langle \rho^{\frac14} \rangle_{\bm x}}{\bar T},\label{ardef}
\end{align}
where $\langle X \rangle_{\bm  x}$ represents a spacial average defined as 
\begin{align}
\langle X \rangle_{\bm x}=\int d^3\vect{x}'~W_{r_T}(\bm x')X(\bm x +\bm
x') \label{sosika}
\end{align}
with $W_{r_T}$ being a window function with a radius $r_T$ around $\bm
x$. This spacial averaging procedure effectively represents the
(instantaneous) acoustic reheating, in which we assume that damping and
thermalization happen instantaneously within the same box. The term
$\langle \rho \rangle^{\frac14}_{\bm x}$ represents the average
temperature when the energy density of photons around ${\bm x}$ would be
damped perfectly and thermalized homogeneously within the radius
$r_T$. Another term $\langle \rho^{\frac14} \rangle_{\bm x}$ represents
the average temperature when the (fluctuating) photon temperature would
be homogenized within the radius $r_T$. From the energy conservation,
this difference yields the temperature rise. One can easily see that
this effect comes from nonlinear relation between the energy density and
the temperature of photons, and is second order in perturbations of the
photon temperature. In fact, inserting the definition of the photon
perturbation (\ref{Tdef}) into (\ref{ardef}) yields $\Delta(\bm x) = 3
\langle \Theta_i^2 \rangle_{\bm x}/2$. Here $\Theta_i$ represents the
temperature perturbation before acoustic reheating happens. 
Realistic
acoustic reheating proceeds through two processes, damping of the photon
perturbations due to Silk damping followed by (homogeneously)
thermalization through the Compton and the double-Compton
processes. Therefore, the actual temperature rise for a conformal time
$\eta$ can be written as
\begin{align}
 \Delta(\eta,\vect x) = \frac32 \langle \Theta_i^2 - \Theta^2(\eta) \rangle_{\vect x}.
\end{align}
Here, we separate two processes explicitly. $\Theta_i^2 - \Theta^2(\eta)$
represents only the damping effect and includes the time evolution of
damping scales. The spacial average in this equation represents only
(homogeneous) thermalization effect and does not include the damping
effect. Then, this equation can be recast into the evolution equation of
$\Delta$ as
\begin{align}
\frac{d \Delta}{d\eta} = -\frac32 \frac{d \langle \Theta^2\rangle_{\bm x}}{d\eta}.\label{delta:eq}
\end{align}
Strictly speaking, thermalization (diffusion) radius $r_T$ also depends
on the conformal time. However, as long as we are interested in the final
temperature rise, we have only to take $r_T$ to be the largest
thermalization radius during acoustic reheating. Therefore, we fix
$r_T$ to be such a radius here and hereafter.

During $5\times 10^4 < z <2\times 10^6$, the double Compton process is
negligible while the Compton process is still efficient.  Then, the
system is locally in kinetic equilibrium, but its spectrum obeys the
Bose-Einstein distribution with a nonzero chemical potential
$\mu$~\cite{Danese:1982,1991A&A...246...49B,Chluba:2006kg}, which
modifies the expression of the temperature rise $\Delta$ for a given
perturbed system of photons, assuming acoustic reheating happens
perfectly and instantaneously, as
\begin{align}
\Delta =\frac{\langle \rho \rangle^{\frac14}_{\bm x}-\langle
 \rho^{\frac14} \rangle_{\bm x}}{\bar T}
- \frac{45\zeta(3)}{2\pi^4}\langle \mu \rangle_{\bm x},
\end{align}
with $\zeta$ being a Riemann zeta function.  Subtraction of the chemical
potential means that the temperature of the Bose-Einstein system is not
the forth root of the energy density. Then, the relation between the
temperature rise, and the photon perturbation is also modified
as~\cite{Chluba:2012gq}
\begin{align}
\frac{d \Delta}{d\eta} \simeq - 2.277 \frac{d \langle \Theta^2\rangle_{\bm x}}{d\eta}.
\label{delta:eq2}
\end{align}
Thus, in order to estimate the amount of the acoustic reheating,
we have only to evaluate how much the temperature perturbations
originated from the primordial curvature perturbations decay due to 
Silk damping. Once the acoustic reheating is taken into account, the
local temperature is modified from (\ref{Tdef}) into
\begin{align}
T(\eta,\vect x,\hat{\vect n}) = \overline{T}(\eta) \bigl[ 1 + \Theta(\eta,\vect x,\hat{\vect n}) + \Delta(\eta,\vect x) \bigr].\label{Tdef2}
\end{align}
It should be noticed that $\Delta$ consists of only a monopole
component because of the thermalization effects
and that it will include the homogeneous part for the entire Universe
 in general, which can reheat the whole Universe.\\

The time evolution of the linear temperature perturbations in the
conformal Newtonian gauge is described by the Boltzmann
equation~\cite{Bond:1984fp,Kodama:1986fg,Ma:1995ey,Kosowsky:1994cy},
\begin{align}
\frac{d}{d\eta} \Theta 
 =\dot{\phi}
 - (\vect{\hat{k}} \cdot \vect{\hat{n}}) \psi +\dot \tau \left[\Theta-\Theta_0+\frac12P_2(\vect{\hat{k}} \cdot \vect{\hat{n}})(\Theta_2+\Theta^P_2+\Theta^P_0)-(\vect{\hat{k}} \cdot \vect{\hat{n}})v\right],\label{boltzmann1}
\end{align}
where $\tau$ is the optical depth, and $\vect{\hat k}\cdot \vect{ \hat
n}$ is a cosine between unit vectors of the photon momentum and the
Fourier momentum.  Over-dots represent partial derivatives with respect to
the conformal time.  Each quantity is Fourier transformed, and $\Theta$
is also expanded by Legendre polynomials $P_l$ as
$\Theta=\sum_l(-i)^l(2l+1)P_l(\hat k\cdot \hat n)\Theta_l$.  $\Theta^P$
is the polarization, and its Legendre coefficients are given in the same
manner.  The definitions of the velocity potential $v$, gravitational
potential $\psi$ and the curvature perturbation $\phi$ are based
on~\cite{Ma:1995ey}.  The third term on the r.h.s. is negligible on
large scales.  Then, the monopole and the dipole components of
temperature perturbations are enhanced due to the Sachs-Wolfe effects
just after the horizon entry.  Once $\phi$ and $\psi$ decay, the
monopole and the dipole oscillate in the inverted phase.  The leading
term proportional to $\dot \tau$ comes from the quadrupole component,
and emerging of such nonzero anisotropic stress induces the diffusion
damping. Below the diffusion scale, the Boltzmann hierarchies are solved
approximately~\cite{Hu:1995en},
\begin{align}
\Theta_1\simeq
 -\frac{1}{\sqrt{3}}\sin(kr_s)\exp\left(-\frac{k^2}{k_D^2}\right) \mathcal R_{\bm
k},\label{Ffgamma1:anal}
\end{align}
where $\mathcal R_{\bm k}$ is primordial curvature perturbations on
comoving slice, $k_D^{-1}$ corresponds to the diffusion scale, and $r_s$
is the sound horizon.  In the quasi tight coupling regime, the following
relations approximately hold, $\Theta^P_2+\Theta^P_0=\frac32 \Theta_2$,
$\Theta_2= 8k\Theta_1/(15\dot \tau)$ and $\partial_\eta
k^{-2}_D=-8/(45\dot \tau)$.  As pointed in~\cite{Jeong:2014gna}, the
diffusion scale should be divided into at least $5$ regions depending on
redshift dependence of $k_D$ and the chemical potential. Taking the
beginning of the hot universe to be $z_0=10^{30}$, in which anisotropic
shear comes from relativistic species such as $\gamma$, $W^{\pm}$ and
$Z$ bosons until $z_1\sim 1.5\times 10^{14}$, damping scale is written
as $k_D\simeq 5.5\times 10^9(1+z)^{0.51}{\rm Mpc}^{-1}$.  Once the weak
bosons acquire their masses, neutrino is a dominant component to smooth
inhomogeneities before neutrino decoupling around $z_2\sim 5.5\times
10^9$, and the diffusion scale is given by $k_D\simeq 5.0\times
10^{-22}(1+z)^{2.7}{\rm Mpc}^{-1}$.  Since neutrino can smooth
inhomogeneities over the whole horizon scale just before the decoupling,
the diffusion scale is almost the same with that at decoupling. Then,
the diffusion scale is almost constant until the photon diffusion scale
goes beyond $k_D \simeq 10^5{\rm Mpc}^{-1}$ around $z_3\sim 8.4\times
10^6$.  $z_5\simeq 5\times 10^4$ is the lower bound since kinetic
equilibrium is not established anymore.  Thus, during $z_5<z<z_3$, the
damping scale is induced by photon diffusion and is written as
$k_D\simeq4.1\times 10^{-6}(1+z)^{3/2}{\rm Mpc}^{-1}$. Though the
diffusion scale is unchanged, we need to distinguish a chemical
potential era from a blackbody era, whose transition happens around
$z_4\simeq 2\times 10^6$.  Combining (\ref{delta:eq}),
(\ref{boltzmann1}), and (\ref{Ffgamma1:anal}), the temperature rise in
the Newton gauge due to the acoustic reheating at the last scattering
conformal time $\eta_{\ast}$ is evaluated as
\begin{align}
\Delta(\eta_{\ast},\vect{x})=&2\int \frac{d^3\vect{k_1}}{(2\pi)^3}
 \frac{d^3\vect{k_2}}{(2\pi)^3}
e^{i (\bm k_1 + \bm k_2)\cdot \bm x}\mathcal W_{r_T}(k)5P_2(\vect{\hat{k}}_1 \cdot \vect{\hat{n}})P_2(\vect{\hat{k}}_2 \cdot \vect{\hat{n}}) 
\langle \sin(k_1 r_s)\sin(k_2 r_s)\rangle_p
\notag \\
&\times\sum^5_{n=1} \alpha_n\left[\exp{\left(-\frac{k_1^2+k_2^2}{k^2_D(z)}\right)}\right]^{z_ {n-1}}_{z_{n}}\mathcal R_{\bm  k_1}\mathcal R_{\bm  k_2},\label{delta:real}
\end{align}
where $\langle \cdots \rangle_p$ is periodic average over the duration
longer than the oscillation period.  $z_n$ and $\alpha_n$ is given by
$(z_0,\cdots,z_5)=(10^{30},1.5 \times 10^{14},5.5\times 10^9,8.4\times
10^6,2\times 10^6,5\times 10^4)$ and
$(\alpha_1,\cdots,\alpha_4,\alpha_5)=(3/2,\cdots,3/2,2.28)$,
respectively. In (\ref{delta:real}), we drop the dipole contribution
proportional to $\Theta_1\dot \Theta_1P_1(\hat k_1)P_1(\hat k_2)$
because its periodic average vanishes, and omitted the higher order
multipoles proportional to $P_l P_{l'}$ with $l \ne l'$, which vanish
thanks to the orthogonality of the Legendre polynomials when we estimate
the angular power spectrum later.  Coarse graining is operated through
the window function $\mathcal W_{r_T}(k)$, which is the Fourier
transformation of $W_{r_T}(\bm x)$ defined in (\ref{sosika}). The
typical scale of such homogenization is given by the maximum scale of
thermalization around $z\simeq 5\times 10^4$. Then, inhomogeneities in
acoustic reheating are erased on scales below $\mathcal O(0.1{\rm
Mpc})$.

In Fourier space, (\ref{delta:real}) can be written as
\begin{align}
\Delta(\eta_{\ast},\vect{k})=& 2 
 \sum^5_{n=1}\alpha_n
 \int\frac{d^3 \vect{k}_1}{(2\pi)^3} \int\frac{d^3 \vect{k}_2}{(2\pi)^3}
 5P_2(\vect{\hat{k}}_1 \cdot \vect{\hat{n}})P_2(\vect{\hat{k}}_2 \cdot \vect{\hat{n}})(2\pi)^3 \delta^{(3)}(\vect{k}-\vect{k}_1-\vect{k}_2)\notag \\
&\mathcal W_{r_T}(k)\langle\sin(k_1 r_s)\sin(k_2
 r_s)\rangle_p\left[\exp\left(-\frac{k_1^2+k_2^2}{k_D^2(z)}\right)\right]^{z_{n-1}}_{z_n}\mathcal
 R_{\bm k_1}\mathcal R_{\bm k_2}.\label{delta:fourier}
\end{align}
Here, it should be noticed that the temperature rise due to
acoustic reheating is smoothed over the diffusion radius $r_T$, so that
the contribution for $k \gg r_T^{-1}$ is significantly
suppressed. However, this fact implies that only the sum of $\vect{k}_1$
and $\vect{k}_2$ must be small and that $k_1$ and $k_2$ themselves can
be large because the acoustic reheating is the second order
effect. This is the essential reason why the perturbations of acoustic
reheating can probe the information of primordial curvature
perturbations on small scales.

Before the end of this section, we shall make a comment on the other
$2$nd order effects.  In the above $\Theta$ is assumed to be given by
the solution of the linearized Boltzmann equation, (\ref{boltzmann1}).
As studied in~\cite{Bartolo:2006cu,Bartolo:2006fj,Huang:2012ub,Pettinari:2013he,Huang:2013qua},
the non-linearity of the Boltzmann and the Einstein equations also
induces non-linear temperature fluctuations of the CMB, namely the
non-linear part of $\Theta$.  While such $2$nd order effects are so far
evaluated around and after the recombination, we investigate a part of
such $2$nd order effects generated well before the recombination in this
paper.  Then the effect estimated in this paper will be separately
treated from the standard $2$nd order effects simply because the
physical origin is different.  Of course, in order to compare with the
observed temperature anisotropies of the CMB, one should take into
account the both effects, which are indistinguishable in fact. Here, to
emphasize and note the importance of $2$nd order effects well before the
recombination, we shall especially focus on this part of specific $2$nd
order effects.

\section{Angular power spectrum}\label{section3}

Temperature anisotropies which we observe are
decomposed using the spherical Harmonics $Y_{lm}$, and their
coefficients are defined as
\begin{align}
a^\Theta_{lm}&=\int d \vect{\hat{n}} Y^*_{lm}(\vect{\hat{n}})\Theta(\bm  x=0, \vect{\hat{n}}),\label{defatheta}\\
a^{\Delta}_{lm}&=\int d \vect{\hat{n}} Y^*_{lm}(\vect{\hat{n}})\Delta(\bm  x=0, \vect{\hat{n}})\label{defadelta},
\end{align}
where we have set the coordinate of an observer to the origin without
loss of generality. $\Theta$ and $\Delta$ above are evolved from the
last scattering surface and hence $\Delta$ has $\hat{\vect n}$
dependence as well in contrast to (\ref{delta:real}). The coefficient of
the total temperature perturbations is given by
$a^T_{lm}=a^\Theta_{lm}+a^{\Delta}_{lm}$. Then, the CMB temperature
angular correlation is written as
$C^{TT}_l=C^{\Theta\Theta}_l+2C^{\Theta \Delta }_l+C^{\Delta \Delta}_l$,
where we have defined
\begin{align}
C^{XY}_{l}=\frac{1}{2l +1}\sum_{m=-l}^l\langle a^{X*}_{lm}a^{Y}_{lm}\rangle,\label{cldef}
\end{align}
and $\langle \cdots \rangle$ represents the ensemble average.  From the
observations of the CMB anisotropies, $C_l^{TT}$ at large scales are
known to be the order of $\mathcal O(10^{-10})$. Therefore, unless each
contribution cancels out accidentally, it should be smaller than
$\mathcal O(10^{-10})$, that is, $C^{\Theta \Delta }_l \lesssim \mathcal
O(10^{-10})$ and $C^{\Delta \Delta }_l \lesssim \mathcal
O(10^{-10})$. These conditions yield new constraints on the primordial
curvature perturbations at small scales. This is the central topic of
this paper.

Using the Fourier modes, we can express (\ref{defatheta}) and
(\ref{defadelta}) as
\begin{align}
a^\Theta_{lm}&=
4\pi(-i)^l\int \frac{d^3 \vect{k}}{(2\pi)^3}Y^*_{lm}(\vect{\hat{k}}){\cal T}_l(k,\eta_0)\Theta(\eta_{\ast},\vect{k}),\label{a:theta}
\\
a^{\Delta}_{lm}&=
4\pi(-i)^l\int \frac{d^3 \vect{k}}{(2\pi)^3}Y^*_{lm}(\vect{\hat{k}}){\cal
 T}_l(k,\eta_0)\Delta(\eta_\ast,\vect{k}),\label{a:delta}
\end{align}
where $\eta_0$ is the present conformal time. The transfer function is
given by ${\cal T}_l \simeq
[\Theta_0(\eta_*)+\Psi(\eta_*)]j_l[k(\eta_0-\eta_*)]\simeq-3j_l(k\eta_0)/5$.
We can use the same transfer function for $\Theta$ and $\Delta$
 because they are indistinguishable.
 Using a relation between spherical harmonics and Legendre polynomials
\begin{align}
\sum^l_{m=-l}Y^*_{lm}(\vect{\hat{k}}_1)Y_{lm}(\vect{\hat{k}}_2)=\frac{2l+1}{4\pi}P_l(\vect{\hat{k}}_1\cdot \vect{\hat{k}}_2),
\end{align}
 (\ref{cldef}) is reduced to
\begin{align}
C^{XY}_{l}=\frac{36\pi}{25}\int\frac{d^3 {\vect k}}{(2\pi)^3}~j^2_l(k\eta_0)P_{XY}(k)
=\frac{36\pi}{25}\int d\ln k ~j^2_l(k\eta_0) {\cal P}_{XY}(k).\label{clxy}
\end{align}
Here $P_{XY}$ and ${\cal P}_{XY}$ are dimensionful and dimensionless
power spectra for $\Theta_{\vect k} = \Theta(\eta_{\ast},\vect{k})$ and
$\Delta_{\vect k} = \Delta(\eta_{\ast},\vect{k})$ and are defined as
\begin{align}
\langle X_{\bm k}Y^*_{\bm k'}\rangle =(2\pi)^3\delta^{(3)}(\bm k-\bm k')P_{XY}(k), 
\quad {\cal P}_{XY}(k) = \frac{k^3}{2\pi^2} P_{XY}(k), 
\end{align}
where $X$, $Y$ is either $\Theta_{\bm k}$ or $\Delta_{\bm k}$.

\section{Power spectrum}\label{section4}

In this section, we estimate the cross- and auto-correlations,
 namely $\langle \Theta_{\bm k}\Delta_{\bm k'}\rangle$ and
 $\langle \Delta_{\bm k}\Delta_{\bm k'}\rangle$, to
 calculate the non-linear corrections to the temperature angular
 power spectrum such as $C^{\Theta \Delta }_l$ and
 $C^{\Delta \Delta }_l$ in the next section.

\subsection{$\Theta $-$\Delta$ cross power spectrum}

First of all, (\ref{delta:fourier}) yields the following cross correlation function:
\begin{align}
\langle \Theta_{\bm  k} \Delta^*_{\bm  k'}
 \rangle=&-\frac{2}{3}\sum^5_{n=1}\alpha_n\int\frac{d^3\vect{k}_1}{(2\pi)^3}\int\frac{d^3\vect{k}_2}{(2\pi)^3}(2\pi)^3\delta^{(3)}(\bm
 k' +\bm k_1 + \bm k_2)5P_2(\vect{\hat{k}}_1 \cdot \vect{\hat{n}})P_2(\vect{\hat{k}}_2 \cdot \vect{\hat{n}})\notag \\
&\times \mathcal W_{r_T}(k)\langle\sin(k_1 r_s)\sin(k_2 r_s)\rangle_p\left[\exp\left(-\frac{k_1^2+k_2^2}{k_D^2(z)}\right)\right]^{z_{n-1}}_{z_n}\langle \mathcal R_{\bm k}\mathcal R_{\bm k_1}\mathcal R_{\bm k_2}\rangle.
\label{crossofdelta}
\end{align}
In this paper, we concentrate on local type non-Gaussianities because
large- and small-scale perturbations must be
 correlated to probe small scale inhomogeneities
 through acoustic reheating on small scales, by using the CMB anisotropies on large
scales. Then, we shall replace the bispectrum of primordial curvature
perturbations $\langle {\cal R}^3 \rangle$ by the scale dependent
non-linear parameter $f_{\rm NL}$ with the aid of (\ref{a1}) and
(\ref{a2}) in the Appendix:
\begin{align}
&{\cal P}_{\Theta\Delta}(k)\simeq \frac{4}{5}\mathcal W_{r_T}(k) x^{n_s-1}{\cal P}_{\cal
 R}^2(k_0)  \sum^5_{n=1}\alpha_n\left[\int d\ln k_1~f_{\rm NL}(k_1,k_1,k) x_1^{n_s-1}\exp{\left(-\frac{2k_1^2}{k^2_D(z)}\right)}\right]^{z_{n-1}}_{z_n},\label{power:rdelta}
\end{align}

where we have assumed $k \ll k_1$ because we are interested in the
corrections coming from small-scale perturbations to the large-scale CMB
anisotropies. Therefore, this expression is exact only in the limit of
$k \rightarrow 0$. For large $k$ modes, the additional suppression
factor $\exp{(-k^2/k_D^2)}$ appears. Here and hereafter, we assumed
$n_s < 4$ otherwise stated. $x$ and $x_1$ are defined by $x = k/k_0$ and
$x_1 = k_1 / k_0$ respectively where $k_0$ represents a wave number at
the pivot scale.  Below we shall consider two simple but interesting
types of scale-dependent non-Gaussianities, which are encoded in the
scale dependence of $f_{\rm NL}$ parameters as discussed in the
appendix: the power law type and the top hat type.

Let us consider the power law type given in (\ref{powerlawfnl:g}) and
(\ref{powerlawfnl:a}). For this type, the scale dependence is defined by
geometric and arithmetic averaged wavenumber, whose cross power spectra
are given by
\begin{align}
{\cal P}^{G/A}_{\Theta \Delta}(k)\simeq\frac25f^{G/A}_{\rm NL}(k) \mathcal
 W_{r_T}(k) x^{n_s-1} {\cal P}_{\mathcal R}^2(k_0)
 \Gamma\left(\epsilon^{G/A}\right)\sum^5_{n=1}\alpha_n\left[\left(\frac{k_D^2(z)}{2k_0^2}\right)^{\epsilon^{G/A}}\right]^{z_{n-1}}_{z_n},\label{calp:g/a}
\end{align}
where $G/A$ indicates geometric average and arithmetic average
respectively, and $\Gamma (x)$ represents the gamma function.  For the former case, $f^G_{\rm NL}(k)=f^{\rm CMB}_{\rm
NL}x^{n_f/3}$ and $\epsilon^G=(n_s+2n_f/3-1)/2$ while, for the latter
case, $f^A_{\rm NL}(k)=f^{\rm CMB}_{\rm NL}(2/3)^{n_f}$ and
$\epsilon^A=(n_s+n_f-1)/2$. Bluer spectrum index induces larger
temperature rise, and this power spectrum is almost scale-invariant on
large scales because their spectral indices are given
 by $n_s + n_f/3 - 1 $ and $n_s-1$ respectively.
 
Next, let us consider the top hat type.  Substituting the top hat type
$f_{\rm NL}$ defined by (\ref{tophatfnl}) into (\ref{power:rdelta}), we
obtain
\begin{align}
{\cal P}^{\rm tophat}_{\Theta \Delta}(k)=& \frac25 \mathcal
W_{r_T}(k) {\cal P}^2_{\mathcal R}(k_0) x^{n_s-1}\sum^5_{n=1}\alpha_n \left[
f_{\rm NL}^{\rm CMB}~\left(\frac{k^2_D}{2k_0^2}\right)^{\frac{n_s-1}{2}} \Gamma\left(\frac{n_s-1}{2}\right)
\right. \notag \\
& \left.
+\widetilde{f}_{\rm NL} 
\left\{ 
x_i^{n_s-1} E_{\frac12(3-n_s)}\left(\frac{2k_i^2}{k_D^2(z)}\right)-
x_f^{n_s-1} E_{\frac12(3-n_s)}\left(\frac{2k_f^2}{k_D^2(z)}\right)
\right\}
\right]_{z_n}^{z_{n-1}},\label{calp:th}
\end{align}
where ${\rm E}_n(x)$ is the exponential integral function defined as
\begin{align}
{\rm E}_n(x)=\int^\infty_1 dt~\frac{e^{-xt}}{t^n}.
\end{align}
The first term represents the contribution coming from the
scale-invariant part $f^{\rm CMB}_{\rm NL}$. By taking into the present
constraint on the (scale-independent) local type $f_{\rm NL}$, this term
is easily shown to be negligible.
This power spectrum is also almost scale-invariant on large scales
because its spectral index is given by $n_s-1$, and we can see that the
amplitudes are proportional to the logarithmic width of the top hat
given by $\log k_i/k_f$ since $E_1(x)\sim \log x$. This implies that
even if we take $k^{-1}_i$ as small as the horizon at the beginning of
the universe, the total contributions to acoustic reheating are not
significant but at most the logarithm of the fraction of $k_f$ and
$k_i$.

\subsection{$\Delta$-$\Delta$ power spectrum}

In this subsection, the power spectrum of $\Delta$ is computed, and  there are 3 types of contributions: two
non-Gaussian connected parts and Gaussian disconnected part.  Therefore,
${\cal P}_{\Delta\Delta}(k)$ is nonzero even if the primordial curvature
perturbations are purely Gaussian.  Using  (\ref{delta:fourier}),
the auto-correlation function $\langle \Delta_{\bm k}\Delta^*_{\bm k'}\rangle$
 is written as
\begin{align}
\langle \Delta_{\bm  k}\Delta^*_{\bm  k'}\rangle=&4\sum_{n,m}\alpha_n\alpha_m\int\frac{d^3\vect{k}_1}{(2\pi)^{3}}\int\frac{d^3\vect{k}_2}{(2\pi)^{3}}\int\frac{d^3\vect{k}_3}{(2\pi)^{3}}\int\frac{d^3\vect{k}_4}{(2\pi)^{3}}\notag \\
&\times 5P_2(\vect{\hat{k}}_1 \cdot \vect{\hat{n}})P_2(\vect{\hat{k}}_2 \cdot \vect{\hat{n}})5P_2(\vect{\hat{k}}_3 \cdot \vect{\hat{n}})P_2(\vect{\hat{k}}_4 \cdot \vect{\hat{n}})
\notag \\
&\times 
(2\pi)^3\delta^{(3)}(\bm k - \bm k_1 - \bm k_2)
(2\pi)^3\delta^{(3)}(\bm k' +\bm k_3 + \bm k_4)
\notag \\
&\times \mathcal W_{r_T}(k)\langle\sin(k_1 r_s)\sin(k_2 r_s)\rangle_p\left[\exp\left(-\frac{k_1^2+k_2^2}{k_D^2(z)}\right)\right]^{z_{n-1}}_{z_n}\notag \\
&\times \mathcal W_{r_T}(k')\langle\sin(k_3 r_s)\sin(k_4 r_s)\rangle_p\left[\exp\left(-\frac{k_3^2+k_4^2}{k_D^2(z)}\right)\right]^{z_{m-1}}_{z_m}\notag \\
&
\times \langle \mathcal R_{\bm k_1}\mathcal R_{\bm k_2}\mathcal R_{\bm k_3}\mathcal R_{\bm k_4}\rangle\label{dkdk}
\end{align}
Ensemble average of the fourfold product of $\mathcal R$ can be divided
into connected and disconnected parts, and the former can be
parametrized as
\begin{align}
\langle \mathcal R_{\bm k_1}\mathcal R_{\bm k_2}\mathcal R_{\bm k_3}\mathcal R_{\bm k_4}\rangle_c =&(2\pi)^3
\delta^{(3)}(\bm k_1+\bm k_2+\bm k_3+\bm k_4)\notag \\
&\times \bigg[\tau_{\rm NL}(k_1,k_2,k_3,k_4)(P_{\mathcal R}(k_{12})P_{\mathcal R}(k_1)P_{\mathcal R}(k_3) + 11 {\rm perms.})\notag \\
&+\frac{54}{25}g_{\rm NL}(k_1,k_2,k_3,k_4)(P_{\mathcal R}(k_{1})P_{\mathcal R}(k_2)P_{\mathcal R}(k_3) + 3 {\rm perms.})\bigg]\label{4rconne}
\end{align}
On the other hand, using Wick's theorem, the latter is reduced to 
\begin{align}
\langle \mathcal R_{\bm k_1}\mathcal R_{\bm k_2}\mathcal R_{\bm
 k_3}\mathcal R_{\bm k_4}\rangle_{dc}=&(2\pi)^3\delta^{(3)}(\bm k_1+\bm k_2)(2\pi)^3\delta^{(3)}(\bm k_3+\bm k_4)P_{\mathcal R}(k_1)P_{\mathcal R}(k_3)\notag \\
&+(2\pi)^3\delta^{(3)}(\bm k_1-\bm k_3)(2\pi)^3\delta^{(3)}(\bm k_2-\bm k_4)P_{\mathcal R}(k_1)P_{\mathcal R}(k_2)\notag \\
&+(2\pi)^3\delta^{(3)}(\bm k_1-\bm k_4)(2\pi)^3\delta^{(3)}(\bm k_2-\bm k_3)P_{\mathcal R}(k_1)P_{\mathcal R}(k_2)\notag \\
=&(2\pi)^3
\delta^{(3)}(\bm k_1+\bm k_2-\bm k_3-\bm k_4) 
\bigg[
(2\pi)^3\delta^{(3)}(\bm k_3 +\bm k_4)P_{\mathcal R}(k_1)P_{\mathcal R}(k_3)\notag \\
&+(2\pi)^3\delta^{(3)}(\bm k_2 -\bm k_4)P_{\mathcal R}(k_1)P_{\mathcal R}(k_2)+
(2\pi)^3\delta^{(3)}(\bm k_2 -\bm k_3)P_{\mathcal R}(k_1)P_{\mathcal R}(k_2)
\bigg].\label{4rdic}
\end{align}

\subsubsection*{Disconnected part}

Collecting disconnected contributions, (\ref{dkdk}) and (\ref{4rdic}) yield
\begin{align}
{\cal P}^{dc}_{\Delta\Delta}(k)\simeq
&\frac{30}{7}\mathcal W_{r_T}^2(k) \mathcal P^2_{\mathcal R}(k_0)~x^3
\int d\ln k_1~x_1^{2n_s-5}
\left(
\sum_{n}\alpha_n
\left[\exp\left(-\frac{2k_1^2}{k_D^2(z)}\right)\right]^{z_{n-1}}_{z_n}
\right)^2,
\end{align}
where we have omitted the term proportional to $\delta^{(3)}(\vect k)$
because such a term does not contribute to $C^{TT}_l$ except the
monopole, which should be included in offsets.
It should be again
noticed that we have assumed $k \ll k_1$ in the same way as the
$\Theta$-$\Delta$ cross power spectrum. Therefore, this expression is
exact only in the limit of $k \rightarrow 0$. For large $k$ modes, the
additional suppression factor $\exp{(-k^2/k_D^2)}$ appears.
After the momentum integration it yields 
\begin{align}
{\cal P}^{dc}_{\Delta\Delta}(k) \simeq 
& \frac{30}{7}~\mathcal W_{r_T}^2(k) \mathcal P_{\mathcal R}^2(k_0)~x^3~
2^{\frac{5-2n_s}{2}} \left(2^{\frac{3-2n_s}{2}}-1\right)\notag \\
&\times 
\Gamma\left(\frac{2n_s-5}{2}\right) \alpha^2_5
 \left(\frac{k_0^2}{k^2_D(z_5)}\right)^{\frac{5-2n_s}{2}},
\label{discopgd}
\end{align}
where we have used $k_D(z_i) > k_D(z_j)$ for $z_i > z_j$.
If we assume the scale-invariant power spectrum with $\mathcal P_{\mathcal R}(k_0=0.05{\rm Mpc}^{-1})\simeq 2.2\times 10^{-9}$, (\ref{discopgd}) yields  
\begin{align}
\mathcal P^{dc}_{\Delta\Delta}(k\to 0)\simeq 61.7\times \left(\frac{k}{k_D(z_5)}\right)^{3}\mathcal P^2_{\mathcal R}(k_0).
\end{align}
This implies that the spectral index of $\mathcal P^{dc}_{\Delta\Delta}$
on large scales is 4, and the pivot scale is $k_D(z_5)\gg k_0$.
Therefore, very little contributions to the COBE scale temperature
perturbations are expected.

\subsubsection*{Connected part}

Among connected contributions, the most important contribution comes
from the term with $\tau_{\rm NL}$, which is written as
\begin{align}
{\cal P}^{\tau}_{\Delta\Delta}(k) \simeq 
&4{\cal P}_{\mathcal R}(k)\mathcal W^2_{r_T}(k)
\int d\ln k_1 \int d\ln k_2~\tau_{\rm NL}(k_1,k_1,k_2,k_2) 
\mathcal P_{\mathcal R}(k_1) \mathcal P_{\mathcal R}(k_2)
\notag \\ &
\times
\left(\sum \alpha_n   
\left[\exp{\left(-\frac{2 k_1^2}{k^2_D(z)}\right)}\right]_{z_n}^{z_{n-1}}
\right)
\left(\sum \alpha_m
\left[\exp{\left(-\frac{2 k^2_2}{k^2_D(z)}\right)}\right]_{z_m}^{z_{m-1}}
\right),\label{Delta_tau}
\end{align}
where we have used (\ref{dkdk}) and (\ref{4rconne}), and have assumed $k
\ll k_1, k_2$ in the same way.  We shall particularly focus on the
geometric averaged type for simplicity because the expression for
the arithmetic averaged one is a bit redundancy.  For the geometric
averaged type, using (\ref{reftaug}), the power spectrum reduces to
\begin{align}
{\cal P}^{\tau,G}_{\Delta\Delta}(k)=\tau^{\rm CMB}_{\rm NL}\mathcal P^3_{\mathcal R}(k_0)x^{n_s-1}\mathcal W^2_{r_T}(k)\left(\Gamma(\epsilon)\sum^5_{n=1}\alpha_n\left[\left(\frac{k^2_D(z)}{2k_0^2}\right)^{\epsilon}\right]^{z_{n-1}}_{z_n}\right)^2,\label{calpd:g}
\end{align}
where $\epsilon=(n_s+n_\tau/2-1)/2$.  The scale dependence is similar to
that of $\mathcal P^G_{\Theta \Delta}$ and is almost scale invariant on
large scales.
On the other hand, in the case with the top hat type non-Gaussianity given in (\ref{tophatfnl}),
(\ref{Delta_tau}) yields
\begin{align}
{\cal P}^{\tau,{\rm top}}_{\Delta\Delta}(k)=
& \mathcal P^3_{\mathcal R}(k_0)\mathcal W^2_{r_T}(k)
x^{n_s-1}\left[ 
\tau^{\rm CMB}_{\rm NL} \Gamma^2\left(\frac{n_s-1}{2}\right) \left(\sum^5_{n=1}\alpha_n\left[\left(\frac{k^2_D(z)}{2k_0^2}\right)^{\frac{n_s-1}{2}}\right]^{z_{n-1}}_{z_n}\right)^2
\right. \notag \\
& + 2 \widehat{\tau}_{\rm NL}~\Gamma\left(\frac{n_s-1}{2}\right)
\left(\sum^5_{n=1}\alpha_n\left[\left(\frac{k^2_D(z)}{2k_0^2}\right)^{\frac{n_s-1}{2}}\right]^{z_{n-1}}_{z_n}\right)
\notag \\
& \qquad
\times \left( \sum^5_{m=1}\alpha_m 
\left[
x_i^{n_s-1} {\rm E}_{\frac{3-n_s}{2}}\left(\frac{k_i^2}{2k_D^2(z)}\right)
- x_f^{n_s-1} {\rm E}_{\frac{3-n_s}{2}}\left(\frac{k_f^2}{2k_D^2(z)}\right)
\right]^{z_{n-1}}_{z_n}
\right)
\notag \\  
& \left. + \widetilde{\tau}_{\rm NL}
\left( \sum^5_{m=1}\alpha_n 
\left[
x_i^{n_s-1} {\rm E}_{\frac{3-n_s}{2}}\left(\frac{k_i^2}{2k_D^2(z)}\right)
- x_f^{n_s-1} {\rm E}_{\frac{3-n_s}{2}}\left(\frac{k_f^2}{2k_D^2(z)}\right)
\right]^{z_{n-1}}_{z_n}
\right)^2
\right].
\label{calpd:th}
\end{align}
In this case, the scale dependence of $\mathcal P^{\tau,\rm
top}_{\Delta}$ is also similar to that of $\mathcal P^{\rm
tophat}_{\Theta\Delta}$ and is almost scale invariant on large scales.
Finally we shall also show the another contribution from the term with
$g_{\rm NL}$ for the sake of completeness:
\begin{align}
{\cal P}^{g}_{\Delta\Delta}(k) \simeq 
& \frac{216}{25} {\cal P}^3_{\mathcal R}(k_0) \mathcal W^2_{r_T}(k)
\int d\ln k_1 \int d\ln k_2~g_{\rm NL}(k_1,k_1,k_2,k_2) 
x_1^{2(n_s-1)} x_2^{n_s-1} \left( \frac{x}{x_1} \right)^3
\notag \\ &
\times
\left(\sum \alpha_n   
\left[\exp{\left(-\frac{2 k_1^2}{k^2_D(z)}\right)}\right]_{z_n}^{z_{n-1}}
\right)
\left(\sum \alpha_m
\left[\exp{\left(-\frac{2 k^2_2}{k^2_D(z)}\right)}\right]_{z_m}^{z_{m-1}}
\right).\label{Delta_g}
\end{align}
Note that the spectral index of this power spectrum is 4 as in the
case with the disconnected part and extremely blue on large scales.

\section{Non-linear corrections to the temperature anisotropies and
 constraints on non-Gaussianities}\label{section5}

Unfortunately, the scale-independent non-Gaussianities of primordial
curvature perturbations which are now strongly constrained by the
observations of the CMB anisotropies have little contribution to
non-linear temperature corrections.  However, the non-Gaussianities might not be (even approximately)
 scale-invariant over a wide range of scales,
 which implies their magnitude on small scales can be large (or small).
As shown thus far, the non-linear corrections to the temperature
anisotropies from acoustic reheating are sensitive to the local type
scale-dependent non-Gaussianities. Then, in this section, we concretely
estimate the CMB anisotropies coming from such temperature perturbations
and give constraints on small-scale non-Gaussianity of primordial
curvature perturbations. Our purpose is not full analysis but to obtain
the Sachs-Wolfe plateau for the low $l$'s.  Therefore, we can safely set
$\mathcal W_{r_T}(k)=1$ because $k r_T \ll 1$ and omit the exponential
suppression factor due to the photon diffusion hereafter (except the
disconnected and the $g_{\rm NL}$ cases as explained later) since the
spherical Bessel functions only project small $k$ modes onto low $l$
plateau.  

The angular cross power spectra between $\Theta$ and $\Delta$
in the case of power law type non-Gaussianity are obtained from
(\ref{clxy}) and (\ref{calp:g/a}) as
\begin{align}
 C^{\Theta \Delta,G}_l\simeq&\frac{9\pi^2}{125}f^{\rm CMB}_{\rm NL}\mathcal P^2_{\mathcal R}(k_0)
\Gamma(\epsilon^{G})
\sum^5_{n=1}\alpha_n\left[\left(\frac{k^2_D}{2k^2_0}\right)^{\epsilon^{G}}\right]^{z_{n-1}}_{z_n}\mathcal G(l,n_s+n_f/3),
\nonumber \\
 C^{\Theta \Delta,A}_l\simeq&\frac{9\pi^2}{125} \left(\frac23\right)^{n_f}
f^{\rm CMB}_{\rm NL}\mathcal P^2_{\mathcal R}(k_0)
\Gamma(\epsilon^{A})
\sum^5_{n=1}\alpha_n\left[\left(\frac{k^2_D}{2k^2_0}\right)^{\epsilon^{A}}\right]^{z_{n-1}}_{z_n}\mathcal G(l,n_s),
\end{align}
where we define
\begin{align}
\mathcal G(l,n)=\left(\frac{k_0\eta_0}{2}\right)^{1-n}\frac{\Gamma\left(l+\frac{n}{2}-\frac12\right)\Gamma\left(3-n\right)}{\Gamma\left(l-\frac{n}{2}+\frac{5}{2}\right)\Gamma^2\left(2-\frac{n}{2}\right)}.
\end{align}
Here we have used the expressions of the power spectra only for
small $k$ modes, which is justified for our purpose because the
spherical Bessel functions only project small $k$ modes onto low $l$
plateau.  For the top hat case, (\ref{clxy}) and (\ref{calp:th}) yield
\begin{align}
C^{\Theta \Delta,{\rm tophat}}_l\simeq&\frac{9\pi^2}{125}\widetilde{f}_{\rm NL}\mathcal P^2_{\mathcal R}(k_0)\mathcal G(l,n_s)\notag \\
&\times \sum^5_{n=1}\alpha_n \left[  
x_i^{n_s-1} E_{\frac12(3-n_s)}\left(\frac{2k_i^2}{k_D^2(z)}\right)-
x_f^{n_s-1} E_{\frac12(3-n_s)}\left(\frac{2k_f^2}{k_D^2(z)}\right)
\right]_{z_n}^{z_{n-1}}.
\end{align}
Angular auto power spectra of $\Delta$ with different scale-dependent
non-Gaussianities are also calculated from (\ref{clxy}), (\ref{calpd:g})
and (\ref{calpd:th}).  The formulas are summarized as
\begin{align}
C^{\Delta \Delta,G}_l\simeq&\frac{9\pi^2}{50}\tau^{\rm CMB}_{\rm NL}\mathcal P^3_{\mathcal R}(k_0)\mathcal G(l,n_s)
\left(
\Gamma(\epsilon)
\sum^5_{n=1}\alpha_n\left[\left(\frac{k^2_D}{2k_0^2}\right)^{\epsilon}\right]^{z_{n-1}}_{z_n}
\right)^2
,
\end{align}
and
\begin{align}
C^{\Delta \Delta,{\rm tophat}}_l\simeq&\frac{9\pi^2}{50} \mathcal P^3_{\mathcal R}(k_0)\mathcal G(l,n_s)\notag \\
&\times
\left[
 2 \widehat{\tau}_{\rm NL}~\Gamma\left(\frac{n_s-1}{2}\right)
\left(\sum^5_{n=1}\alpha_n\left[\left(\frac{k^2_D(z)}{2k_0^2}\right)^{\frac{n_s-1}{2}}\right]^{z_{n-1}}_{z_n}\right) \right.
\notag \\
& \qquad
\times \left( \sum^5_{m=1}\alpha_m 
\left[
x_i^{n_s-1} {\rm E}_{\frac{3-n_s}{2}}\left(\frac{k_i^2}{2k_D^2(z)}\right)
- x_f^{n_s-1} {\rm E}_{\frac{3-n_s}{2}}\left(\frac{k_f^2}{2k_D^2(z)}\right)
\right]^{z_{n-1}}_{z_n}
\right)
\notag \\
& \left.
\quad + \widetilde \tau_{\rm NL} \left(\sum^5_{n=1}\alpha_n \left[  
x_i^{n_s-1} E_{\frac12(3-n_s)}\left(\frac{2k_i^2}{k_D^2(z)}\right)-
x_f^{n_s-1} E_{\frac12(3-n_s)}\left(\frac{2k_f^2}{k_D^2(z)}\right)
\right]_{z_n}^{z_{n-1}}\right)^2 \right],
\end{align}
where we have omitted the first term in (\ref{calpd:th}) because
it is negligible. 

\begin{figure}
\begin{center}
\includegraphics[width=7.5cm]{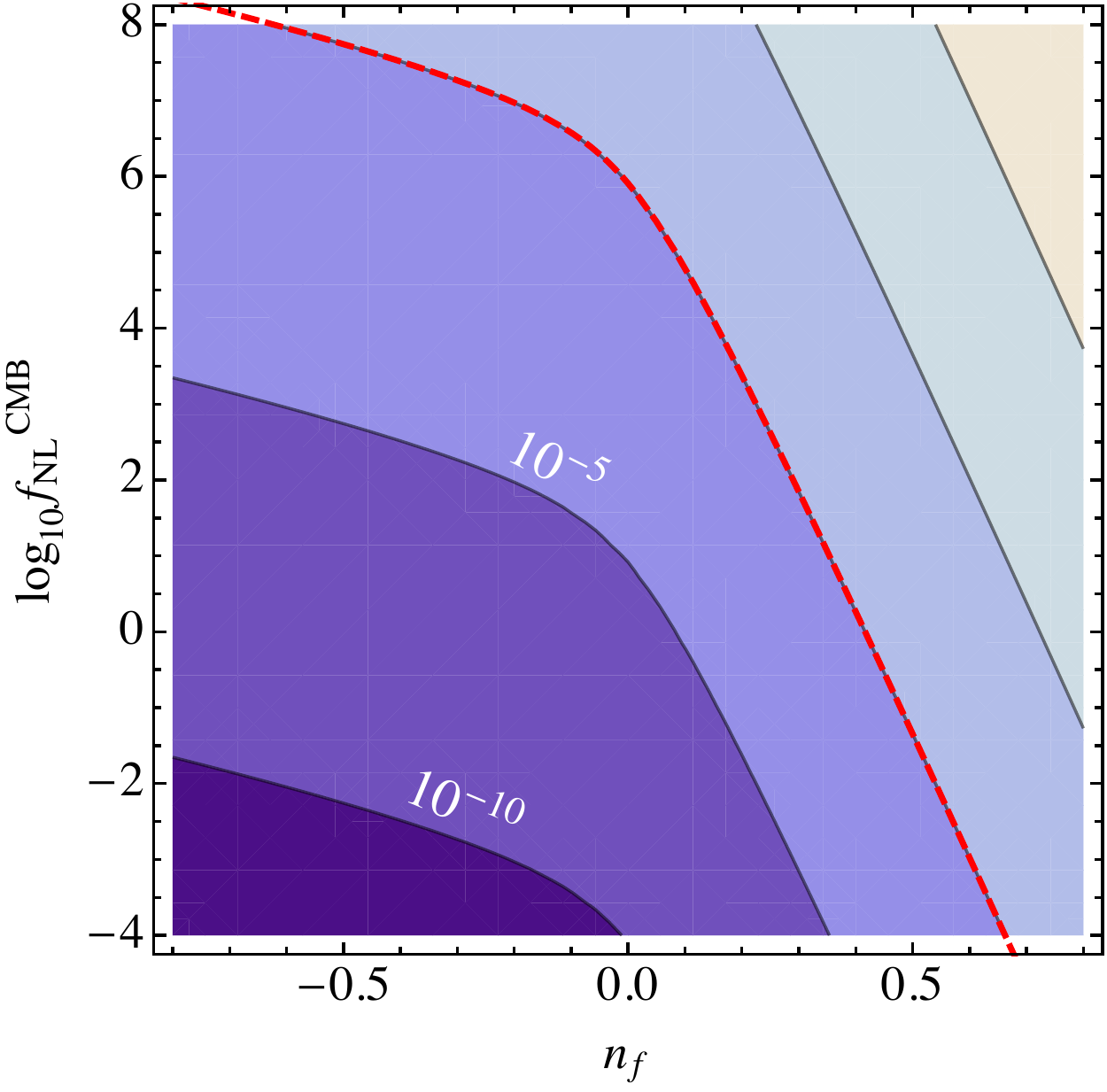}~~~~\includegraphics[width=7.5cm]{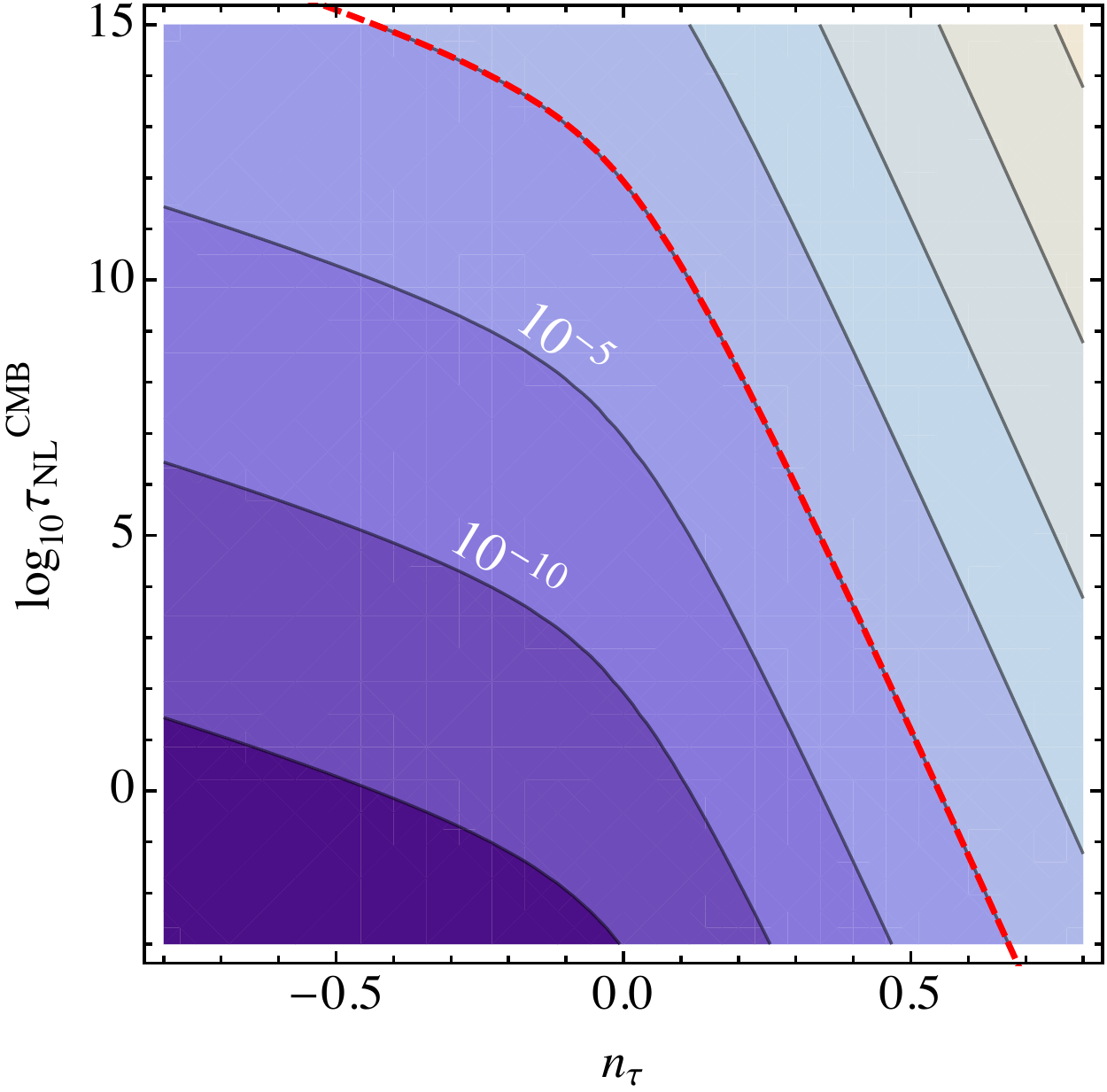}~~~~
\end{center}
\caption{The constraints on $n_f$-$f^{\rm CMB}_{\rm NL}$ plane (left)
and $n_\tau$-$\tau^{\rm CMB}_{\rm NL}$ plane (right).  The interval of
the contours of the $2C^{\Theta\Delta}_{50}/C^{\Theta\Theta}_{50}$(left)
and $C^{\Delta\Delta}_{50}/C^{\Theta\Theta}_{50}$(right) are $10^5$, and
red dashed lines correspond to the lines of
$2C^{\Theta\Delta}_{50}=C^{\Delta\Delta}_{50}=C^{\Theta\Theta}_{50}$. Only
the region below the red line is allowed.}  
\label{fig_c12}
\end{figure}

Now that all the formulae are derived, let us
investigate the physical implications of the results. As we have already
stated before, given the fact that the linear theory prediction well
explains the observations of the CMB anisotropies, the corrections
obtained above should not exceed the observed values and hence must be
subdominant. From this condition, we obtain the constraints on
small-scale non-Gaussianities. In Fig.~\ref{fig_c12}, we show the
constraints on the non-Gaussianities with the power law type of
the geometrical average at pivot scale $k_0=0.05 {\rm Mpc}^{-1}$ as a
function of each spectrum index with fixing $n_s = 1$. Horizontal axes
represent the spectrum indexes of the non-Gaussianities, and the
vertical axes represent the logarithms of the non-linear parameters.
The contours of the
2$C^{\Theta\Delta}_{50}/C^{\Theta\Theta}_{50}$($C^{\Delta\Delta}_{50}/C^{\Theta\Theta}_{50}$)
are inserted at the intervals of $10^5$ in the left(right) panel, and
red dashed lines correspond to the lines along which the non-linear
corrections are comparable with $C^{\Theta\Theta}_{50}$. Then, only
the region below the red line is allowed.

\begin{figure}
\centering
\includegraphics[width=8cm]{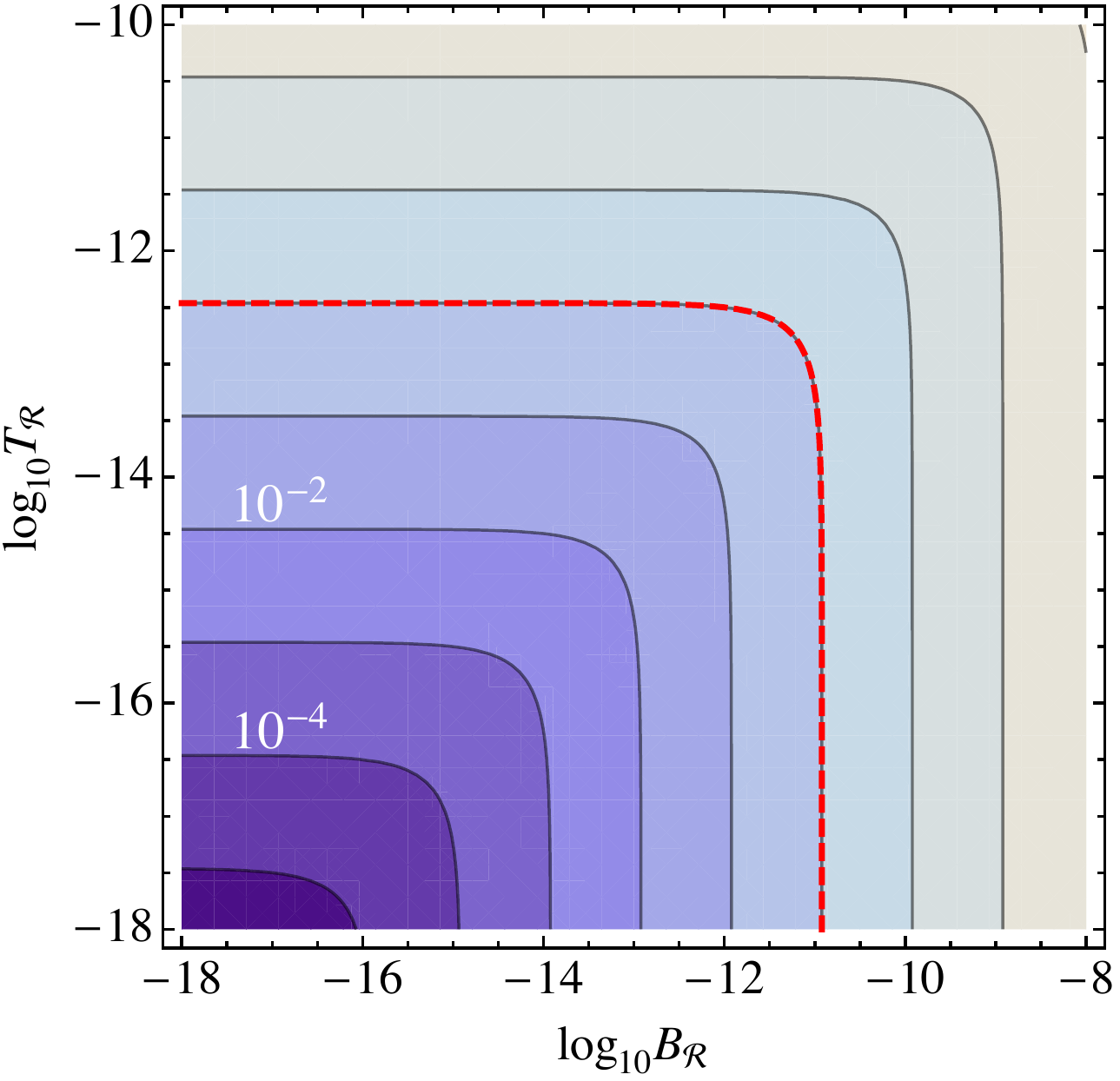}
\caption{The figure shows the constraints on $B_{\mathcal R}=\widetilde
f_{\rm NL}\mathcal P^2_{\mathcal R}(k_0)$ and trispectrum $T_{\mathcal
R}=\widetilde \tau_{\rm NL}\mathcal P^3_{\mathcal R}(k_0)$. The contours
of
$(2C^{\Theta\Delta}_{50}+C^{\Delta\Delta}_{50})/C^{\Theta\Theta}_{50}$
are inserted at the interval of $10$ in the case with
$\log_{10}k_i/k_f=4$, and a red dashed line is the line along which the
non-linear correction is equal to $C^{\Theta\Theta}_{50}$. Only the
region below the red line is allowed.
}  \label{fig_c3}
\end{figure}

In fig.~\ref{fig_c3}, we show the constraints on the top hat type
non-Gaussianities in the case with $n_s=1$.  For simplicity, we also set
$\widehat{\tau}_{\rm NL} = 0$. Instead of giving the constraints on
$\widetilde{f}_{\rm NL}$ and $\widetilde{\tau}_{\rm NL}$, we give the
constraints directly on the magnitudes of bispectrum $B_{\mathcal
R}=\widetilde f_{\rm NL}\mathcal P^2_{\mathcal R}(k_0)$ and trispectrum
$T_{\mathcal R}=\widetilde \tau_{\rm NL}\mathcal P^3_{\mathcal R}(k_0)$.
This is because $\widetilde{f}_{\rm NL}$ and $\widetilde{\tau}_{\rm NL}$
are normalized not by the (unknown) power spectrum at the corresponding
scale but by that at the large (CMB) scale, that is, $P_{\mathcal
R}(k_0)$. The figure shows the constraints on bispectrum $B_{\mathcal
R}$ and trispectrum $T_{\mathcal R}$. The contours of
$(2C^{\Theta\Delta}_{50}+C^{\Delta\Delta}_{50})/C^{\Theta\Theta}_{50}$
are inserted at the interval of $10$ in the case with
$\log_{10}k_i/k_f=4$, and a red dashed line corresponds to the line
along which the non-linear correction is equal to
$C^{\Theta\Theta}_{50}$. Then, only the region below the red line is
allowed.
  
The disconnected part and the non-linear term proportional to $g_{\rm
NL}$ do not contribute significantly. This is because such
configurations do not pick up a pair of perturbations which are confined
in the diffusion scale. 
The non-linear corrections to angular power spectra coming from these
parts become
\begin{align}
C^{\Delta\Delta,dc/g}_{l}=&\frac{36\pi}{25}\int\frac{d^3 k}{(2\pi)^3}j^2_l(k\eta_0)P^{dc/g}_{\Delta}(k)\notag \\
=&\frac{36\pi}{25(2\pi^2)}P^{dc/g}_{\Delta}(k\to 0)\int dk k^2j_l^2(k\eta_0)\mathcal W_{r_T}(k).
\end{align}
Note that, in these cases, the use of the power spectra only for small
$k$ modes is not justified because their spectra indexes are 4 and
extremely blue, as already pointed out in (\ref{discopgd}) and
(\ref{Delta_g}). Therefore, we need to take into account the window
function $\mathcal W_{r_T}$ or the exponential suppression factor
$\exp{(-k^2/k_D^2)}$ for large $k$ modes. Since $r_T \sim
k_D(z_5)^{-1}$, here, we simply take the Gaussian window function with
$r_T$, $W_{r_T}(\vect x) = (1/(\sqrt{2\pi r_T})^3) \exp(-r^2/(2 r_T^2))$
and its Fourier component ${\cal W}_{r_T}(\vect k) = \exp(-r_T^2
k^2/2)$, into account as a suppression factor for large $k$ modes, which
enables us to integrate these expressions explicitly,
\begin{align}
C^{\Delta\Delta,dc/g}_{l}=&\frac{9}{25}\frac{1}{\sqrt{2\pi}\eta_0^2r_T}P^{dc/g}_{\Delta}(k\to 0)\left[1-\frac{r_T^2}{2\eta_0^2}l(l+1)+\mathcal O \left(\frac{r_T^4}{\eta_0^4}\right)\right]\notag \\
\simeq&10^{-7}\frac{P^{dc/g}_{\Delta}(k\to 0)}{(2\pi^2)} ,\label{cdddg}
\end{align}
where we have used 
\begin{align}
\int dk k^2j_l^2(k\eta_0)\exp\left(-\frac12 r_T^2 k^2 \right)=&\frac\pi 2\frac{1}{\eta_0 r_T^2}\exp\left(-\frac{\eta_0^2}{r_T^2}\right)I_{l+\frac12}\left(\frac{\eta_0^2}{r_T^2}\right)\notag \\
=&\frac{\sqrt{\pi}}{2\sqrt{2}\eta_0^2r_T}\left[1-\frac{r_T^2}{2\eta_0^2}l(l+1)+\mathcal O\left(\frac{r^4_T}{\eta_0^4}\right)\right],
\end{align}
and the asymptotic expression for the modified Bessel function
$I_{l+\frac12}(z)$ given by
\begin{align}
I_{l+\frac12}(z)=\frac{e^z}{\sqrt{2\pi z}}\left[1-\frac{l(l+1)}{2z}+\mathcal O(z^{-2})\right].
\end{align}
Note that the spectral indexes are 4, which implies that the
dimensionful power spectra $P^{dc/g}_{\Delta}(k\to 0)$ for small $k$
modes have no $k$ dependence. We have also substituted
$r_T\simeq\mathcal O(0.1{\rm Mpc})$.  

In passing, one also finds that (\ref{discopgd}) and
(\ref{cdddg}) yield the weak constraints on power spectrum at $k\simeq
k_{D}(z_5)$
\begin{align}
\mathcal P_{\mathcal R}(\mathcal O(10{\rm Mpc}^{-1}))\lesssim \mathcal O(0.1).
\end{align}
However, we have already known the stronger constraints on the scale by
spectral
distortions~\cite{Hu:1994bz,1991MNRAS.248...52B,1991ApJ...371...14D,Chluba:2012we,Khatri:2012rt,Khatri:2012tv,Chluba:2012gq}.
On the other hand, the constraints on $g_{\rm NL}$ can be about $10^7$
times weaker than those for $\tau_{\rm NL}$, comparing the cases with scale-independent local type $\tau_{\rm NL}$ and $g_{\rm NL}$. 

\section{Conclusions and discussions}

We have discussed the inhomogeneities of the acoustic
reheating developing the previous works which have mainly focused on
 the homogeneous part. Such inhomogeneities arise
from the higher order correlation functions of primordial curvature
perturbations because the acoustic reheating is
a non-linear phenomenon, which first appear at the second order in perturbation.
Produced (secondary) temperature perturbations
are, in general, indistinguishable from the standard temperature
perturbations linearly coming from primordial curvature perturbations.
Given the fact that the linear theory prediction well matches with the
CMB observations, the non-linear corrections should be subdominant
compared to the observational $C^{TT}_l$, i.e. the order of $10^{-10}$
at large scales. Based on this condition, we gave constraints on higher
order correlation functions, especially, the scale dependent local type
non-Gaussianities of primordial curvature perturbations, in which large- and small-scale perturbations are correlated.
These configurations of bispectra of primordial curvature perturbations are complementary to those directly inferred from the three point correlation functions of the CMB anisotropies.
Though the power spectra of the CMB anisotropies are used to constrain
the (small-scale) higher order correlation functions of primordial
curvature perturbations, the extension to the higher order
correlation functions such as bispectrum and trispectrum of the CMB
anisotropies to further constrain primordial curvature perturbations
is straightforward.

In this paper, we have used only the total temperature perturbations to
constrain the small-scale non-Gaussianities. However, the acoustic
reheating can generate isocurvarture perturbations because the ratio of
the number density of photon and baryon (dark matter, neutrino) is
perturbed universally, as already pointed out in
\cite{Jeong:2014gna}. The constraint on isocurvature perturbations
may give more stringent constraints on the small-scale primordial
perturbations. This topic will be discussed elsewhere \cite{OY}.

\acknowledgments 

We would like to thank Jens Chluba for careful reading our manuscript and useful comments.
We also would like to thank Teruaki Suyama for helpful comments. 
A.~N would like to thank Cyril Pitrou and Atsushi Taruya for fruitful
discussions.  A.~N would also like to thank the Yukawa Institute for
Theoretical Physics at Kyoto University for the hospitality during his
stay when part of this work was done. A.~N. is supported by Grant-in-Aid
for JSPS Fellows No. 26-3409 and M.~Y. is supported by the JSPS
Grant-in-Aid for Scientific Research Nos. 25287054 and 26610062.

\begin{appendix}
\section{Scale-dependent non-Gaussianity}

In Fourier space, the bispectrum of the comoving curvature perturbation
is written as
\begin{align}
\langle \mathcal R_{\bm  k_1}\mathcal R_{\bm  k_2}\mathcal R_{\bm  k_3}\rangle =(2\pi)^3\delta^{(3)}(\bm  k_1 +\bm  k_2 + \bm  k_3)B_{\mathcal R}(k_1,k_2,k_3),\label{a1}
\end{align}
and the local type one is parameterized as follows.
\begin{align}
B_{\mathcal R}(k_1,k_2,k_3)=-\frac65f_{\rm NL}(k_1,k_2,k_3)P^2_{\mathcal R}(k_0)(x_1^{n_s-4}x_2^{n_s-4} + 2 {\rm perms.}),\label{a2}
\end{align}
where $x_\alpha=k_\alpha/k_0$. Now, let us consider scale-dependent
local type non-Gaussianity defined as~\cite{Sefusatti:2009xu}
\begin{align}
f^{G/A}_{\rm NL}(k_1,k_2,k_3)=f^{\rm CMB}_{\rm NL}X_{G/A}^{n_f}
\end{align}
with $X_G=(x_1x_2x_3)^{\frac13}$ and $X_A=(x_2+x_2+x_3)/3$.  $f^{\rm
CMB}_{\rm NL}$ is the $f_{\rm NL}$ parameter at the pivot scale $k_0$.
In our cases, we can pick up the configuration of squeezed isosceles
triangles, $k \ll k_1$, so we obtain the following forms.
\begin{align}
f^G_{\rm NL}(k_1,k_1,k)&= f^{\rm CMB}_{\rm NL}x_1^{\frac{2n_f}{3}}x^{\frac{n_f}{3}}\label{powerlawfnl:g},\\
f^A_{\rm NL}(k_1,k_1,k)&\simeq f^{\rm CMB}_{\rm NL}\left(\frac{2x_1}{3}\right)^{n_f}\label{powerlawfnl:a}.
\end{align}
For the top hat type, using the top hat function defined as
$W(x)=\theta(x-x_i)-\theta(x-x_f)$ with $\theta(x)$ being the step
function, $f^{\rm tophat}_{\rm NL}(k_1,k_2,k_3)$ can be written as
\begin{align}
f^{\rm tophat}_{\rm NL}(k_1,k_2,k_3)=&f^{\rm CMB}_{\rm NL}+\widetilde f^{(1)}_{\rm NL}(W(x_1)+2{\rm perms.})
\notag \\
&+\widetilde f^{(2)}_{\rm NL}\left(W(x_1)W(x_2)+ 2{\rm perms}.\right)\notag \\
&
+\widetilde f^{(3)}_{\rm NL}W(x_1)W(x_2)W(x_3).\label{tophatdef0}
\end{align}
Assuming that the squeezed isosceles triangle configurations are
enhanced, that is $W(x) \simeq 0$, we can simplify (\ref{tophatdef0}) to
\begin{align}
f^{\rm tophat}_{\rm NL}(k_1,k_1,k)=f^{\rm CMB}_{\rm NL}+\widetilde f_{\rm NL}W(x_1),\label{tophatfnl}
\end{align}
where $\widetilde f_{\rm NL}=2\widetilde f^{(1)}_{\rm NL}+\widetilde
f^{(2)}_{\rm NL}$. The trispectrum is also introduced in the same
manner with the bispectrum, and we do not repeat it
here~\cite{Byrnes:2006vq}.  Assuming the power law type, the scale
dependence can be written as
\begin{align}
\tau_{\rm NL}(x_1,x_2,x'_1,x'_2)=\tau^{\rm CMB}_{\rm NL}Y^{n_\tau}_{G/A},
\end{align}
where we have defined $Y_G=(x_1x_2x_3x_4)^{\frac14}$ and
$Y_A=(x_1+x_2+x_3+x_4)/4$.  For the geometric averaged type, the above
formula can be reduced to
\begin{align}
T^G_{\mathcal R}(k_1,k_1,k'_1,k'_1)=\tau^{\rm CMB}_{\rm NL}x_1^{n_\tau/2}{x'_1}^{n_\tau/2}.\label{reftaug}
\end{align}
Top hat type is also defined as,
\begin{align}
\tau^{\rm tophat}_{\rm NL}(x_1,x_1,x'_1,x'_1)=\tau^{\rm CMB}_{\rm NL}
+\widehat \tau_{\rm NL} \left( W(x_1) + W(x'_1) \right)+\widetilde \tau_{\rm NL}W(x_1)W(x'_1).\label{tophatfnl}
\end{align}

\end{appendix}

\end{document}